\documentclass[sigconf]{acmart}

\AtBeginDocument{%
	\providecommand\BibTeX{{%
			Bib\TeX}}}

\copyrightyear{2024}
\acmYear{2024}
\setcopyright{rightsretained}
\acmConference[MSR '24]{21st International Conference on Mining Software Repositories}{April 15--16, 2024}{Lisbon, Portugal}
\acmBooktitle{21st International Conference on Mining Software Repositories (MSR '24), April 15--16, 2024, Lisbon, Portugal}\acmDOI{10.1145/3643991.3644918}
\acmISBN{979-8-4007-0587-8/24/04}

\usepackage{amsmath,amsfonts}
\usepackage{algorithmic}
\usepackage{graphicx}
\usepackage{textcomp}
\usepackage{xcolor}
\usepackage{tabularx,colortbl}
\usepackage{multirow}

\usepackage{etex}
\usepackage{booktabs}
\usepackage[utf8]{inputenc}
\usepackage[ruled,vlined]{algorithm2e}
\usepackage[T1]{fontenc}
\usepackage{microtype}
\usepackage{graphicx}
\usepackage{paralist}
\usepackage{tabularx}
\usepackage{balance}
\usepackage{multicol}
\usepackage{multirow}
\usepackage{pbox}
\usepackage{enumitem}
\usepackage{lscape}
\usepackage{pifont}
\usepackage{xspace}
\usepackage{url}
\usepackage{tikz}
\usepackage{float}
\usepackage[TABBOTCAP]{subfigure}
\usepackage{ragged2e}
\usepackage{fontawesome}
\usepackage[figuresright]{rotating}
\usepackage{bbding} % checkbox

\usepackage{adjustbox}

\newcommand{\ie}{\emph{i.e.,}\xspace}
\newcommand{\eg}{\emph{e.g.,}\xspace}
\newcommand{\etc}{etc.\xspace}
\newcommand{\etal}{\emph{et~al.}\xspace}
\newcommand{\secref}[1]{Section~\ref{#1}\xspace}

\newcommand{\figref}[1]{Fig.~\ref{#1}\xspace}

\newcommand{\tabref}[1]{Table~\ref{#1}\xspace}

\newcommand{\cg}{ChatGPT\xspace}

\newboolean{showcomments}

\setboolean{showcomments}{true}

\ifthenelse{\boolean{showcomments}}
{\newcommand{\nb}[2]{
		\fbox{\bfseries\sffamily\scriptsize#1}
		{\sf\small$\blacktriangleright$\textit{#2}$\blacktriangleleft$}
	}
}
{\newcommand{\nb}[2]{}
}

\newcommand{\manually}{1,501\xspace}
\newcommand{\validInstances}{467\xspace}
\newcommand{\conflicts}{380\xspace}
\newcommand{\prs}{159\xspace}
\newcommand{\issues}{143\xspace}
\newcommand{\commits}{165\xspace}

\newcommand{\categories}{45\xspace}
\newcommand{\categoriesAll}{52\xspace}

\def\BibTeX{{\rm B\kern-.05em{\sc i\kern-.025em b}\kern-.08em
		T\kern-.1667em\lower.7ex\hbox{E}\kern-.125emX}}
\begin{document}
	
	\title{Unveiling ChatGPT's Usage in Open Source Projects:\\A Mining-based Study}
	
	\author{Rosalia Tufano}
	\authornote{Both authors contributed equally to the paper.}
\affiliation{%
  \institution{\small{SEART @ Software Institute\\Universit\`a della Svizzera italiana}}
 \country{Switzerland}
}
\author{Antonio Mastropaolo}
\authornotemark[1]
\affiliation{%
  \institution{\small{SEART @ Software Institute\\Universit\`a della Svizzera italiana}}
 \country{Switzerland}
}
\author{Federica Pepe}
\affiliation{%
  \institution{\small{Department of Engineering\\University of Sannio}}
  \country{Italy}
}
\author{Ozren Dabi\'c}
\affiliation{%
  \institution{\small{SEART @ Software Institute\\Universit\`a della Svizzera italiana}}
  \country{Switzerland}
}
\author{Massimiliano Di Penta}
\affiliation{%
  \institution{\small{Department of Engineering\\University of Sannio}}
 \country{Italy}
}
\author{Gabriele Bavota}
\affiliation{%
  \institution{\small{SEART @ Software Institute\\Universit\`a della Svizzera italiana}}
  \country{Switzerland}
}
	
	\renewcommand{\shortauthors}{Tufano \etal}

	\begin{abstract}
		Large Language Models (LLMs) have gained significant attention in the software engineering community. Nowadays developers have the possibility to exploit these models through industrial-grade tools providing a handy interface toward LLMs, such as OpenAI's ChatGPT. While the potential of LLMs in assisting developers across several tasks has been documented in the literature, there is a lack of empirical evidence mapping the actual usage of LLMs in software projects. In this work, we aim at filling such a  gap. First, we  mine \manually commits, pull requests (PRs), and issues from open-source projects by matching regular expressions likely to indicate the usage of ChatGPT to accomplish the task. Then, we manually analyze these instances, discarding false positives (\ie instances in which ChatGPT was mentioned but not actually used) and categorizing the task automated in the \validInstances true positive instances (\commits commits, \prs PRs, \issues issues). This resulted in a taxonomy of \categories tasks which developers automate via ChatGPT. The taxonomy, accompanied with representative examples, provides (i) developers with valuable insights on how to exploit LLMs in their workflow and (ii) researchers with a clear overview of tasks that, according to developers, could benefit from automated solutions. %In addition, we inspected all \comments reviewers' comments posted in the \prs PRs (partially) automated via ChatGPT, classifying the type of feedback provided on the automated task (\eg identification of functional bugs, suggestions for refactoring). Such an analysis partially discloses shortcomings in the usage of LLMs for software-related tasks, pointing to areas in need for further research.
	\end{abstract}

	\begin{CCSXML}
		<ccs2012>
		<concept>
		<concept_id>10011007.10011006</concept_id>
		<concept_desc>Software and its engineering~Software notations and tools</concept_desc>
		<concept_significance>500</concept_significance>
		</concept>
		</ccs2012>
	\end{CCSXML}
	
	\ccsdesc[500]{Software and its engineering~Software tools}

	\keywords{ChatGPT, Empirical study}
	
	\maketitle

	% !TEX root = main.tex
%%%%%%%%%%%%%%%%%%%%%%%%%%%%%%%%%%%%%%%%
%%%%%%%%%%%%%%%%%%%%%%%%%%%%%%%%%%%%%%%%
\section{Introduction} \label{sec:intro}
%%%%%%%%%%%%%%%%%%%%%%%%%%%%%%%%%%%%%%%%
%%%%%%%%%%%%%%%%%%%%%%%%%%%%%%%%%%%%%%%%

Recommender systems for software engineers have been defined by Robillard \etal \cite{RobillardWZ10} as:

\begin{quote}
	\emph{``Software applications that provide information items estimated to be valuable for a software engineering task in a given context''}
\end{quote}

Over the years, researchers have developed various forms of recommenders, aimed at suggesting relevant code elements for a given task \cite{nguyen:icse2012,allamanis:fse2015,wen:icse2021}, helping to fix bugs \cite{xia:tse2017,tufano:tosem2019,bader:oopsla2019,mashhadi:msr2021} and vulnerabilities \cite{han:icsme2017,Gao:tosem2021,Chen:tse2023}, and even to automatically document software systems \cite{sridhara:icse2011,moreno:icse2015,aghaj:tse2019}.

In the last decade, the increasing gain of maturity and improvement of deep learning architectures, the availability of hardware infrastructures, and of data from forges such as GitHub has opened the road towards the development of recommenders able not only to better solve the aforementioned problems, but also to perform tasks for which no recommender was previously thought in the past, including generating entire code blocks \cite{svyatkovskiy:fse2020,ciniselli:tse2021}, automatically reviewing source code \cite{tufano:icse2021,tufano:icse2022,li:esecfse2022,thongtanunam:icse2022}, or generating scenarios for automatically reproducing issues \cite{ZhaoSLZWKHY22,FazziniMBWOP23,BernalCardenasCHMCPM23}.

The advent of Large Language Models (LLMs) and, lately, of LLM-based  chat bots such as \cg \cite{chatgpt} has opened new development landscapes. In such a context, a developer can, from a single tool, receive help for a wide number of tasks: one can ask \cg to design an architecture, write or complete source code to achieve a given task, review and possibly refactor/optimize existing code, repair bugs, generate tests, and so on. In other words, today's ChatGPT and tomorrow's similar tools from other providers will gradually become the main source of help for developers, essentially replacing what a colleague, user manuals, the Web (including forums such as Stack Overflows) and, recently, code completion specialized tools such as GitHub Copilot \cite{copilot}, have done so far.

Given such a scenario, it would be worthwhile understanding how developers have leveraged \cg  so far to achieve different goals. Certainly, this goal could possibly be achieved through interviews and survey questionnaires. However, we have decided to follow a radically different approach, \ie by mining traces of \cg usages in GitHub:  commit messages, issues, and pull requests (PRs). 

\eject

This is because on the one hand, we could observe how developers ``admit'' the use of ChatGPT in their (open-source) projects, but, also, how such code is being reviewed before being merged. 

To conduct our study, we first mined all commits, issues, and PRs from GitHub that match the keyword ``\cg". Then, we extracted n-grams surrounding the word \cg and manually reviewed them for further filtering. This allowed us to filter the initial sample to reduce the chance of false positives, \eg an issue mentioning \cg but not using it for the automation of a task. Then, we performed an open coding based on card sorting \cite{spencer2009card} on all \manually candidate instances (\ie commit/issue/PR) we identified, classifying the \cg purpose for each instance (\ie why it was used) or discarding the instance as a false positive.  

As a result, we obtained a taxonomy of purposes for using \cg in the automation of a software-related task. The taxonomy features seven root categories and \categoriesAll categories in total.

For each category, we discuss typical use cases, as well as implications for practitioners and researchers. Also, we highlight and discuss scenarios for which the use of \cg turned out to be failing, counterproductive, or risky for the given activity.

All data used in our study is publicly available \cite{replication}.
	% !TEX root = main.tex
%%%%%%%%%%%%%%%%%%%%%%%%%%%%%%%%%%%%%%%%
%%%%%%%%%%%%%%%%%%%%%%%%%%%%%%%%%%%%%%%%
\section{Study Design} \label{sec:design}
%%%%%%%%%%%%%%%%%%%%%%%%%%%%%%%%%%%%%%%%
%%%%%%%%%%%%%%%%%%%%%%%%%%%%%%%%%%%%%%%%

The \emph{goal} of the study is to unveil the purposes for which LLM recommenders are used to support the development of open-source projects. The \emph{context} consists of \cg, as a representative of state-of-the-art LLMs, and of \manually manually inspected commits, PRs, and issues sampled from open source projects hosted on GitHub. Our study aims at answering the following research question:

\begin{quote}
\emph{What are the software-related tasks for which developers document the support received by \cg?}
\end{quote}

We answer this research question by mining, from development artifacts, traces of \cg usages. We focus on artifacts for which it is possible to perform keyword-matching queries on GitHub. As such, we search for commit messages, PRs, and issues mentioning \cg in their textual content (\secref{sub:mining}). We do not consider GitHub discussions, as we are interested to analyze text directly traceable to software artifacts. Then, we manually inspect \manually instances with the goal of categorizing the task(s) supported by \cg (if any) in each of them (\eg \emph{generate tests}, \emph{code review}) --- see \secref{sub:collection}. The obtained categories of tasks have then been used to derive a taxonomy of tasks supported by \cg. Such a taxonomy %, shown in \figref{fig:taxonomy}, 
provides (i) developers with a comprehensive catalog of usage scenarios in which LLM recommenders can be leveraged; and (ii) researchers with software-related tasks which could benefit from automation, possibly through specialized solutions to be developed rather than via generic LLM recommenders such as \cg.

In the following, we detail the steps behind our study design.  

\subsection{Mining Candidate Instances}
\label{sub:mining}
The goal of this step is to identify commits, PRs, and issues in which \cg has been \emph{likely} used to support one or more tasks. False positive instances (\eg instances in which \cg was mentioned but not actually leveraged) will be discarded in a later stage. 

We started by querying---on June 12, 2023---the GitHub APIs to identify all commits, PRs, and issues containing the word ``\cg''. For commits, the search was performed on the commit message/body, while for PRs and issues the target was their title and description. The output of this step were 186,425 commits, 15,629 PRs, and 31,934 issues (233,988 overall instances). By inspecting the retrieved instances, we noticed a predominance of false positives, mostly due to projects which integrate \cg (\ie use the \cg APIs) to offer features to their users (\eg a chat bot) rather than using it for automating software-related tasks. 

We then performed a first filtering to automatically discard as many false positives as possible. To this aim, we extracted from the collected instances all forward/backward 2-grams and 3-grams containing the word ``\cg''. For example, let us assume that a commit message features the sentence: ``\emph{Implemented matrix transposition with the help of \cg}''. In this case, we extract the following backward $n$-grams: ``\emph{of \cg}'', and ``\emph{help of \cg}''.  Instead, a PRs titled \emph{Used \cg to implement tests} will result in the backward 2-gram ``\emph{used \cg}'' and in the forward $n$-grams ``\emph{\cg to}'' and ``\emph{\cg to implement}''.

We then sorted all extracted $n$-grams in ascending order of frequency, and inspected those appearing in at least 0.02\% of the instances ($>$1k instances). We classified each $n$-gram as likely indicating the \cg support in a task (\eg ``\emph{\cg to generate}'') or as likely indicating false positives (\eg ``\emph{\cg API integration}''). This resulted in a set of 34 relevant $n$-grams available in our replication package \cite{replication}. We excluded all instances not containing at least one of such $n$-grams, and all those belonging to GitHub repositories having less than 10 stars in an attempt to filter out toy projects. Finally, we removed duplicates (\eg duplicated commits due to forked repositories) obtaining a final set of \manually instances, distributed as follows: 527 commits, 327 PRs, and 647 issues. The \manually manually analyzed issues, commits, and PRs belong to 732 different projects.

\subsection{Manual Analysis and Taxonomy Definition}
\label{sub:collection}
The goal of the manual analysis was to characterize within each of the \manually instances the task(s) (partially) automated using \cg. Five authors (from now on evaluators) were involved in the manual inspection. Each instance has been independently inspected by two evaluators. The whole process was supported by a web app we developed that implemented the required logic and provided a handy interface to categorize the instance. For each instance, the evaluator was presented with: (i) the metadata as returned by the GitHub APIs (\eg for a commit: its author, message, body, date, \etc); (ii) the $n$-gram that was matched in that specific instance (\eg ``\emph{\cg to generate}''); and (iii) the link to the instance on GitHub for an easier inspection. 

The categorization required the assignment of one or more lebels to an instance, describing the automated task(s) (\eg \emph{refactoring code}, \emph{write documentation}). In case the manual inspection revealed that \cg was not actually used to automate software-related tasks, the instance was discarded. 

Since there are no documented taxonomies of software-related tasks automated with the support of LLM recommenders, we followed an open coding strategy \cite{spencer2009card}. 

Specifically,  each evaluator could introduce a new label, as they felt it was needed to properly describe the automated task(s). After the label was added, it became available, through the web app, to the other evaluators. While this goes against the notion of open coding, in a scenario in which there are no pre-defined categories this helps to reduce the chance of multiple evaluators defining similar labels to describe the same task while not introducing a substantial bias in the process. 

It is important to mention that the labeling process has not been performed in a single shot, but rather in three rounds each involving roughly $\frac{1}{3}$ of the instances to inspect. At the end of each round the authors met to revise the set of labels defined up to that moment by (i) renaming unclear labels; (ii) merging similar labels, \ie labels describing the same automated task but with different wordings; and (iii) agreeing on irrelevant labels, actually indicating instances to discard (\ie unrelated to tasks supported by \cg). 

Once all \manually instances have been inspected by two evaluators, we solved conflicts. As for the relevance labeling, we found conflicts in 17\% of the cases, with Cohen's $k=0.64$, which is considered a strong agreement \cite{kappa}. 

For what concerns the (open-coded) categories, we found differences in \conflicts cases ($\sim$25\% of instances). While such a percentage may look high, this can be easily explained by two design choices. 
The first is the already mentioned lack of pre-defined categories. This implies that two evaluators defining semantically equivalent but different labels to describe an automated task (\eg \emph{create tests} \emph{vs} \emph{test writing}) would generate a conflict. The second concerns our conservative definition of conflicts: We considered an instance as a conflict if two evaluators assigned a different set of labels to the instance, \emph{even if the two sets partially overlapped}. Conflicts also arose if one of the two evaluators discarded the instance as a false positive while the other labeled it. Each conflict has been inspected in pairs by two additional evaluators, who discussed and solved it. In the end, \validInstances instances were kept and classified, distributed as follows: \commits commits, \prs PRs, and \issues issues. The 467 classified instances belong to 358 projects, having [min=8, 1Q=60, median=444, 3Q=3,075, max=179,567] stars, and [min=0, 1Q=15, median=77.5, 3Q=535, max=89,252] forks.

The \categories labels defined through the above-described process have been used to build a hierarchical taxonomy of software-related tasks for which \cg provided (partial) automation. Two of the authors created a preliminary version of the taxonomy which has then been refined in two rounds by collecting the feedback of all five authors involved in the labeling. 
	% !TEX root = main.tex
%%%%%%%%%%%%%%%%%%%%%%%%%%%%%%%%%%%%%%%%
%%%%%%%%%%%%%%%%%%%%%%%%%%%%%%%%%%%%%%%%
\section{Results Discussion} \label{sec:results}
%%%%%%%%%%%%%%%%%%%%%%%%%%%%%%%%%%%%%%%%
%%%%%%%%%%%%%%%%%%%%%%%%%%%%%%%%%%%%%%%%

\begin{figure*}
	\centering
	\includegraphics[width=\linewidth]{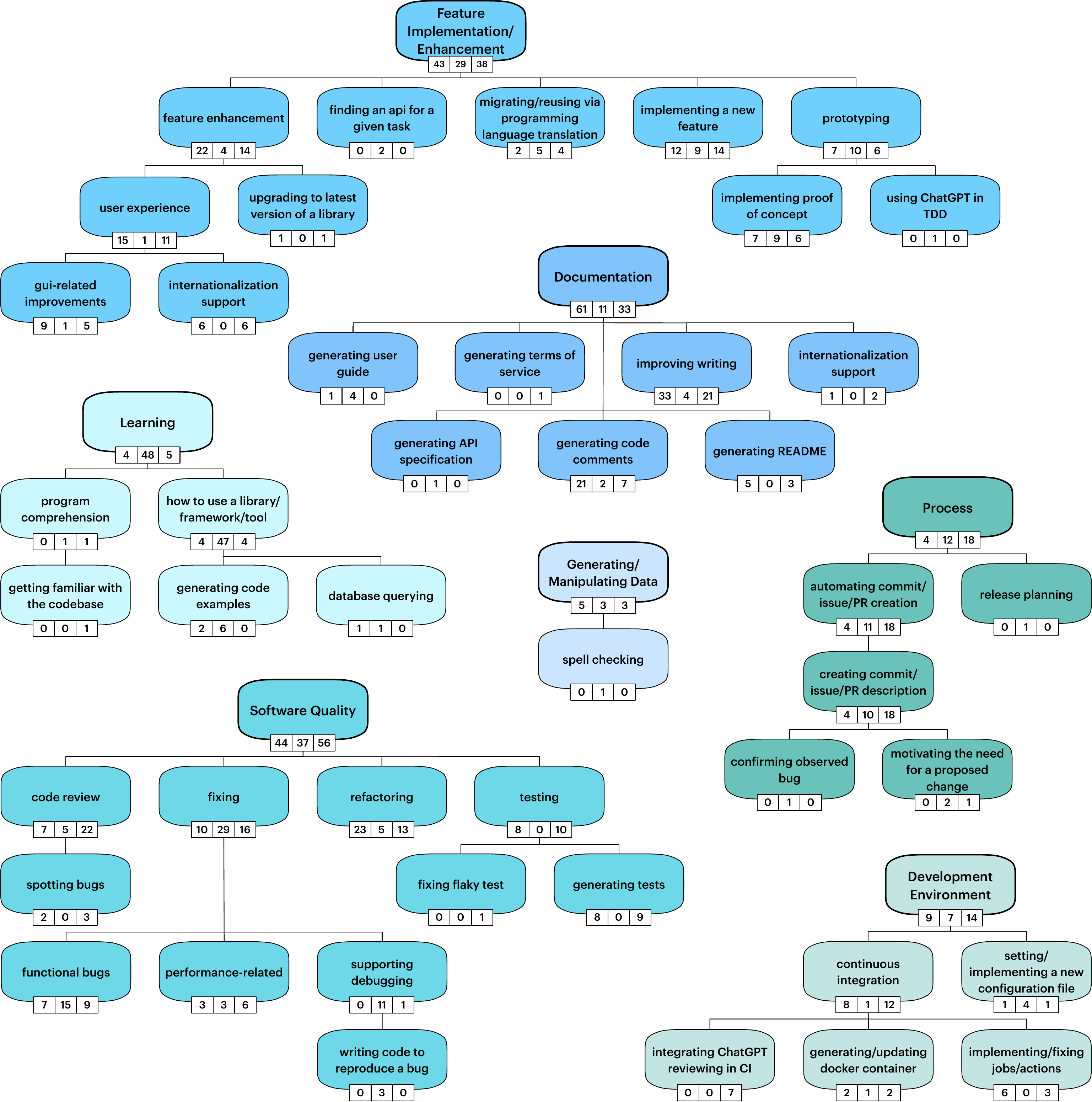}
	\caption{Taxonomy of types of tasks automated via \cg}
	\label{fig:taxonomy}
\end{figure*}

\figref{fig:taxonomy} depicts the taxonomy of tasks automated via \cg. The taxonomy is composed of seven trees, each grouping together related tasks: \emph{feature implementation/enhancement}, \emph{process}, \emph{learning}, \emph{generating/manipulating data}, \emph{development environment}, \emph{software quality}, and \emph{documentation}. 
The numbers attached to each task $T_i$ indicate, from the right to the left, the number of commits, issues, and PRs in which we found evidence of $T_i$'s automation using \cg. For example, we found a total of 110 instances (43 commits, 29 issues, and 38 PRs) in which \cg has been used to automate the implementation or enhancement of a feature. 

Note that the sum of the number of instances in all tasks is greater than the total number of valid instances we inspected (\validInstances), since one instance may have required the support of \cg for multiple tasks, \eg \emph{generating tests} and their related \emph{code comments}. 
Also, note that the number of instances in a parent category is not always the sum of the instances in its child categories. For example, consider the \emph{software quality $\rightarrow$ fixing $\rightarrow$ supporting debugging} category: Such a task has been automated in 12 instances (11 issues and 1 PRs) and has one child category named \emph{writing code to reproduce a bug}, automated in 3 issues. The reason for such a discrepancy is that in 12 instances it was clear that \cg has been used to support the debugging process, but only in three of those cases the classification could be even more precise and refer to the specific task of helping to reproduce a bug.
%
%\begin{figure*}
%	\centering
%	\includegraphics[width=\linewidth]{img/cateogy_definitions.pdf}
%	\caption{Definitions for the categories in our taxonomy}
%	\label{fig:definitions}
%\end{figure*}
%
%In addition, \figref{fig:definitions} provides definitions for each category, helping the reader in better understanding and navigating our taxonomy.

In the following, we discuss the seven main categories of automated tasks by reporting qualitative examples and discussing implications for practitioners (see \faLaptop~icon) and researchers (\faLightbulbO). We also showcase \cg's limitations when used for the automation of the related tasks (\faWarning). Due to the lack of space, we do not discuss all \categoriesAll categories in our taxonomy, but only the main ones. However, our replication package \cite{replication} provides the complete dataset reporting, for each category, the instances assigned to it.

\subsection{Feature implementation/enhancement}
This category features tasks related to the usage of \cg as a support for implementing and enhancing software features. We start by commenting two of its related, but differing, subcategories: \emph{implementing a new feature} and \emph{prototyping}. The former refers to the usage of \cg as a support to implement a specific \emph{part} of a feature that the developer is working on. This means that the developer delegates the implementation of a specific functionality, which is then manually integrated with the rest of the code needed for the feature. An example is the PR \#37233 from the \texttt{woocommerce} project \cite{pr:1328}, in which the PR author states: ``\emph{I wrote a Python script with \cg to parse csv files since we need to update this payment list quarterly}''. 

The latter, instead, refers to the usage of \cg as a way to quickly implement either (i) a complete feature that can be used as a starting prototype for reasoning about the addition of the new feature in the project, or (ii) the whole starting version of a project, on which developers can work and build on top. An example of the first scenario is the PR \#73 from the \texttt{nix-gaming} project \cite{pr:1486} in which the contributor proposes the addition of an autoupdater script commenting ``\emph{I don't know if this is the right way to do it [$\dots$] Credits to \cg for the script}''. While this PR has been approved and merged, it is representative of those instances in which the contributor explicitly states to be unsure about what was accomplished using \cg, or even  declared that they were completely unfamiliar with coding while contributing PRs or issues --- see \eg \cite{issue:958}: ``\emph{I started to edit the files with the support by \cg --- as said, I have no idea about coding}''.  \faLaptop~Similarly to what happens when defining onboarding and contribution guidelines in open source projects \cite{asfcontrib,eclipsecontrib}, it may be desirable to define guidelines about contributing with AI-generated code, \ie a project may decide to only welcome contributions from \cg by users that are confident in assessing the correctness of the generated code. 

Also, projects may need to adapt the code review process, \eg relying less on (semi-) automated code quality check (\eg a linter to check code quality) when AI-generated contributions come from users having little or no programming expertise. 

\faLightbulbO~This finding is also relevant for research, as, for example, it may impact studies involving contributors of OSS (\eg studies on newcomers similar to those of Steinmacher \etal \cite{SteinmacherCTG16} or Zhou and Mockus \cite{ZhouM10}). Related studies in the future should be careful about surveying developers that have only submitted AI-generated contributions, as they may not be representative of the target population of OSS developers. Clearly, the development landscape may also significantly change, as contributors mainly relying on AI when submitting code could become the norm.

\faWarning~In general, there is a clear risk related to the ownership and understanding of code contributed via \cg, especially when it is used to contribute complete features. Such a problem has been well-summarized in a comment of a PR we inspected \cite{pr:1428}: ``\emph{[$\dots$] I want to make something clear about code suggestions done by \cg: deferring to an AI bot is not the same as code ownership [$\dots$] the idea that an author puts some code into a commit and sends it means they should have an intellectual understanding of it. PR authors should own the code they send -- ownership in the sense of being able to advocate for the code}''. 

The consequence is that AI-generated code may require a more thorough quality assurance, but also may lead to issue triaging problems and in general maintenance issues in the absence of a real owner. Concerning the second usage scenario for prototyping (\ie using \cg to draft the whole starting version of a project), a concrete example is the issue \#1 from the \texttt{apple-notes-to-sqlite} project \cite{issue:815} titled ``Initial proof of concept with \cg''. 

The issue documents the conversation between the repository's owner and \cg, and resulted in the implementation of the first prototype of an application exporting Apple notes to SQLite. \faLaptop~This project counts 120 stars on GitHub at the date of writing and is a concrete example of how \cg can provide a jumpstart in software development.  

Still related to prototyping, we also found one issue \cite{issue:880} in which developers discuss the possibility of using ``\emph{\cg to generate reliable code through a semi-automated Test-Driven Development (TDD) process that incorporates feedback loops}'' (\emph{using \cg in TDD} category in our taxonomy). 
\faLaptop~From a practitioner's perspective, this would lead to a different metaphor, in which the developer is mainly in charge of writing tests letting the LLM generate code.

\faLightbulbO~From a research perspective, this requires empirical investigations, as it has been done in the past for conventional TDD (\eg \cite{BaldassarreCFJR21,FucciETOJ17}) thus defining a suitable process with AI in-the-loop. 

Another popular sub-category is the \emph{feature enhancement} task, in which \cg is used to enhance an already existing feature (as opposed to help contributing with a new feature). This includes generic enhancements such as writing CSS for existing web pages \cite{issue:1026} or adding options to a feature \cite{pr:1432}, as well as more specific improvements that can be seen in the category's subtree. For example, we found 12 instances (6 commits + 6 PRs) in which \cg has been used for internationalization purposes \cite{pr:1294}, mostly related to translating elements in the GUI, including error messages. \faWarning~In some of these cases the reviewers asked whether the contributor was actually familiar with the target language or if, instead, they were just running \cg and reporting the translation, with the risk of introducing internationalization issues. \faLightbulbO~Such a finding is relevant for researchers working on detecting and fixing internationalization issues \cite{velasquez-internationalization20}. For example, \cg could produce mistakes different from those typically committed by developers who manually implement internationalization.

The last sub-category we discuss is the one related to \emph{migrating/reusing via programming language translation}, a task supported by \cg in 11 of the inspected instances. \faLaptop~Practitioners used \cg to automatically translate code snippets across languages, allowing for possibilities of reuse that were unimaginable before (\eg reusing code across projects written in different languages). 

An interesting example is the PR \# 4559 from the \texttt{garden} project \cite{pr:1495} in which the contributor used \cg to translate from Javascript to Typescript the code of a third-party project which has not been updated in the last six years and was known to be affected by vulnerabilities. As documented by the contributor: ``\emph{I didn't find an easy fix $\dots$ so I just created a fork of [third-party project] and asked \cg to convert it to Typescript, removed the dependency on [third-party project] and later tweaked it to make sure that everything works}''. \faLightbulbO~Our findings confirm the relevance of research targeting the automated translation of software across programming languages \cite{nguyen:ase2014,nguyen:icse2014}, but it also highlights the very strong performance of what should be considered the state-of-the-practice and, therefore, a baseline for comparing new approaches in this research thread. \cg seems to be able to generalize across several languages, even by translating hundreds of lines of code, see \eg the Javascript to Python translation in the \texttt{textual-paint} project \cite{commit:518}.

\faWarning~At the same time, recent research has pointed out perils of ML-based language translation \cite{malyala2023mlbased}, especially because the translation may not take into account that different programming languages may follow different programming paradigms (\eg object oriented \emph{vs} functional), and the result could just be ``Java with a Python syntax'' or something similar. As Malyala \etal suggest \cite{malyala2023mlbased}, a combination of ML-based translation with static analysis and rule-based translation could be a pragmatic road to follow.

\subsection{Process}
The \emph{process} category groups instances mentioning the usage of \cg to support activities related to the development process, \eg \emph{release planning}, or the automation of steps needed to \emph{create commits, PRs, issues}, \eg  generating a PR description.

We found a single issue in which \cg has been used to come up with ideas on how to improve a software project (\emph{release planning} label in our taxonomy) \cite{issue:577}. While this is a single data point and should be taken as such, we found this usage of \cg extremely interesting, since it goes substantially further than what state-of-the-art tools supporting release planning are able to do. The latter usually mine data (\eg app reviews \cite{CiurumeleaSPG17,PanichellaSGVCG15,ScalabrinoBRPO19}) to help developers summarizing the customers' feedback and come up with aspects to improve in the software. \faLaptop~\cg does not require any sort of data mining on the developer's side and, as visible from the issue we inspected \cite{issue:577}, can be queried for general ideas about what to improve in a software or even on how to improve a specific feature/quality aspect of the software (\eg ``\emph{Can you suggest improvements to make the help system more useful for data scientists?}'', ``\emph{Suggest ways to make the help system more useful for developers}'').  The contributor confirmed that ``\emph{\cg came up with [$\dots$] pretty good ideas}''. 
\faWarning~One clear limitation is that requirement crowdsourcing may require up-to-date sources of information (\eg recent information about the features that competitive software implements) which \cg may not have. Also, practitioners may be afraid to prompt \cg with sensitive, market-competition-related information. \faLightbulbO~To cope with both issues, researchers may develop approaches based on Retrieval Augmentation Generation (RAG) \cite{LewisPPPKGKLYR020} to combine LLMs with private or up-to-date resources.

In 32 instances developers used \cg to automatically generate a commit message (\eg \cite{commit:288}) or a PR/issue description (\eg \cite{issue:716}). The automatic generation of commit messages \cite{Dong22,Liu18,Siyuan17,Wang21} and of PR title/descriptions \cite{FANG2022111160,Liu2019,Irsan2022} has been tackled by several researchers, especially after the wide adoption of deep learning in software engineering. While we are not aware of empirical comparisons performed between these tools and \cg two observations can be made. \faLightbulbO~First, the instances in our dataset show an impressive capability of \cg in summarizing even complex changes spanning several files, while studies in the literature documented strong limitations of techniques for the automated generation of commit messages, mostly targeting the low-hanging fruits (\eg \emph{Add README}) \cite{liu:ase2018}. Second, empirical comparisons should carefully consider which test instances to use, considering that the dataset on which \cg has been trained is not publicly available. While this does not make it possible to ensure a lack of overlap between training and test set, an easy solution is to use very recent commits/PRs/issues as test set, since those are unlikely to have been seen by the model behind \cg.  

Our taxonomy also features two specializations of the \emph{creating commit/PR/issue description} category. The first concerns a scenario in which \cg has been used to confirm an observed failure as a bug: ``\emph{[failure description] asking \cg suggested it to be a bug}'' \cite{issue:626}. The second represents instances in which \cg was used to better motivate a proposed change. An example of this second usage scenario is the issue \#1648 from the \texttt{js-libp2p} project \cite{issue:907}, being a feature request including a question posed to \cg about the usefulness of the proposed feature (``\emph{Can't we already do this with just a libp2p stream? Why do we need HTTP?}'') with the LLM providing four disadvantages of not having such a feature. 

In other cases falling in this category, the contributor just describes a chat they had with \cg that helped them in coming up with the issue/PR (\eg ``\emph{this is a PR to improve the way we store thumbnails in the data folder; after a nice chat with \cg I discovered why most apps do this}'' \cite{pr:1496}). 

\faWarning~Despite the very successful applications of \cg to \emph{process}-related tasks, we observed cases of what has been recently defined as artificial hallucination  \cite{Ji:hallucination}, namely confident responses provided by an AI such as \cg which look plausible to the human interacting with it but that are clearly wrong. This is reflected in negative reactions to suggestions given by \cg about features to improve/implement (\eg ``\emph{I can't really do anything about that, and it would defeat the entire purpose of [project]}'' \cite{issue:1081}).

\subsection{Learning}
The \emph{learning} category tree mostly features issues opened by users of a software project as a result of problems they are experiencing in using the related library/framework/tool. In doing so, they mention their attempt to solve the faced problem by asking \cg (\eg \cite{issue:680,issue:729}). \faWarning~This is a category of task for which \cg showed clear limitations related to the previously mentioned artificial hallucination issue. In 36 out of the 47 opened issues, the indications provided by \cg on how to solve the problem faced by the user were wrong, resulting in negative comments either by the user itself when opening the issue (\eg ``\emph{I even asked \cg who made up some configuration options that do not exist}'' \cite{issue:729}) or by the developers replying to the issue (\eg ``\emph{just stop asking \cg about this thing, because the data that was used to train it only spans until 2021, which means everything created from 2022 (including [project]) onwards is outside of its knowledge domain; If you ask about something it doesn't know, it will make up fake answers that don't work at all, and fake libraries that don't exist}'' \cite{issue:988}).
\faWarning~Similarly to release planning, leveraging outdated knowledge of LLMs can be risky, therefore they need to be complemented with alternative approaches.

The \emph{learning} tree also features two instances in which \cg has been used to understand code. In these two cases, the LLM provided useful support to the developer, even in understanding code automatically generated by a framework by reverse engineering it: ``\emph{This code is very difficult for people to read because it is compiled by webpack. However, \cg completed reverse-engineering in a short time}'' \cite{issue:1151}. \faLaptop~This shows the potential of \cg in supporting program comprehension and on the research side \faLightbulbO~suggest investigations aimed at assessing the impact of \cg in program comprehension, as well as approaches to support the use of third-party LLMs on private code or other software artifacts.

\subsection{Generating/manipulating data}
Developers use \cg to easily \emph{generate/manipulate data}. The variety of data involved in this category includes strings appearing in the UI (\eg ``\emph{add \cg suggestions to bully messages}'' \cite{commit:495}, ``\emph{Add extra motivational messages generated by \cg}'' \cite{commit:367}), fake data needed to fill templates (\eg ``\emph{add more fake data in the sample using \cg}'' \cite{commit:198}), or more intellectual content such as math problems for an educational project (\eg ``\emph{use \cg to generate these problems and create an initial solution}'' \cite{issue:886}). 

\faLightbulbO~\cg seems to be particularly suited in the generation of data for which correctness is not a strong requirement (\eg fake data, augmenting UI-related strings handling a dialog with the user). This makes it a suitable tool to automatically generate test inputs since even implausibly generated inputs could represent a good opportunity to assess the robustness to wrong inputs. 
\faWarning~However, in this scenario, several risks may arise. First, LLMs can be subject to bias \cite{ChakrabortyM0M20,ChakrabortyMM21} and may generate unwanted discriminatory or offensive text. Moreover, it cannot be excluded that LLMs could be subject to adversarial attacks, leading to the generation of unwanted outputs as it has been shown for other recommender systems \cite{NguyenSRPR21}.

\subsection{Development environment}
This tree groups together instances in which \cg has been used to support and (partially) automate activities related to the \emph{development environment}. The most popular application of \cg is its \emph{integration as reviewer in the continuous integration and delivery (CI/CD) pipeline}. In this scenario, \cg is used to comment about contributed code and identify bugs and/or suboptimal implementation choices (\eg ``\emph{added bot reviewer powered with \cg to help us with PR reviews}'' \cite{pr:1481}).
\faLaptop~The \cg-based review is usually integrated in the continuous integration pipeline and aims at providing a first quick feedback to the contributor, without replacing (but supporting) the human reviewer. Furthermore, \cg is usually combined with classic lint tools looking for issues and assessing test coverage. 
 
 \faLightbulbO~Such an application confirms the relevance of the recent line of research related to the automation of code review tasks \cite{tufano:icse2022}, for which \cg could become a baseline for comparison. Future research should also consider how to properly leverage LLM-recommenders such as \cg to obtain code reviews in line with an organization/project's own coding styles and guidelines.
 \faWarning~A clear issue is the need for passing code to \cg, which may not be acceptable (and even forbidden) in industrial environments. In such cases, approaches leveraging local LLMs are to be preferred. 
 
Other tasks automated via \cg concern the \emph{implementation/fixing of jobs/actions} in continuous integration scripts and the \emph{generation/updating of docker containers}. While \cg can be a good aid to draft CI/CD scripts, this is one of those tasks in which the long training time needed for LLMs and, as a consequence, their inability to be continuously retrained to be updated, represents a strong limitation. Indeed, technologies such as GitHub actions which are used to achieve CI/CD are relatively new and rapidly evolving. This resulted in PRs contributing with CI/CD scripts created with the help of \cg which, accordingly to the reviewers, were using outdated actions and commands (\eg ``\emph{the actions used are outdated}'' \cite{pr:1487}). \faWarning~This highlights one strong limitation of LLMs: They might not be suitable in rapidly evolving contexts such as young technologies, programming languages, \etc

\faLightbulbO~In these cases, it is possible that smaller, specialized models that can be quickly retrained might be more suitable and reliable.

\subsection{Software quality}
\emph{Software quality} is the largest tree in our taxonomy in terms of number of instances (137). This indicates that developers largely leverage \cg for automating tasks related to software quality improvement.

\emph{Refactoring} operations recommended by \cg are widely implemented by developers. This may include simple renaming \cite{pr:1192} as well as more complex code transformations such as converting a recursive function into an iterative one: ``\emph{Thanks to \cg for doing the recursion to iterative conversion}'' \cite{commit:366}. \faLightbulbO~While we found a wide variety of refactoring actions automated via \cg, we observed a lack of code transformations involving multiple files, such as extract class refactoring. This is likely due to the limited view that \cg has of the software systems, given its (current) lack of integration in the IDE. For such cases, approaches to generate suitable prompts helping \cg to produce responses for more complex refactoring scenarios may be desirable.

\emph{Functional bugs} have been fixed with the help of \cg (31 instances in our taxonomy). \faWarning~In these instances we observed two of the previously discussed issues affecting the usage of \cg in open-source projects. First, \cg has been used by inexperienced programmers to submit patches, possibly with little understanding of the contributed code: ``\emph{Was getting the error $\dots$ so I used \cg to fix it, not sure how GitHub works $\dots$ so I gonna put it here}'' \cite{issue:628}.
Second, the inability of \cg to cope with complex code, similar to what we inferred looking at the automated refactorings: \eg  ``\emph{Do not trust \cg to fix complex code depending on multiple files; \cg has no idea of the scope nor the current state of the codebase, so it will not be able to give a valid answer $\dots$}'' \cite{issue:628}.

Besides the explicit bug fixes suggested by \cg, the LLM is also used to \emph{support debugging}. \faLaptop~This mostly comes in two fashions. The first is the expected one, with a user observing a failure in a code and asking \cg what was causing it: \eg ``\emph{I asked \cg what this error could be, and the AI gave me an important clue $\dots$}'' \cite{issue:660}. The second is the usage of \cg to generate a minimal, reproducible example \cite{issue:949}. \faLightbulbO~As of today, there are state-of-the-art approaches supporting the automated reproduction of bugs \cite{BernalCardenasCHMCPM23,FazziniMBWOP23,ZhaoSLZWKHY22}. This is mainly possible because such approaches are (i) tailored for specific categories of applications, \eg mobile apps, and (ii) can access the whole application code base. In the future, LLM-based approaches should be therefore able to use such information to support the automated reproduction of bugs. 

The \emph{code review} subtree collects the usage of \cg as a reviewer, mostly for incoming PRs. It is worth commenting on the difference between this subtree and the previously discussed \emph{development environment $\rightarrow$ continuous integration $\rightarrow$ integrating \cg reviewing in CI}. The latter concerns instances in which developers manifested their interest in integrating \cg in the CI/CD pipeline as a reviewer. The former, instead, groups instances in which the outcome of a code review performed by \cg was discussed, even when \cg was not integrated into the workflow but queried through its user interface. 

Similarly, the \emph{code review $\rightarrow$ spotting bugs} differs from \emph{support debugging} since in the former the developers were not aware of the bugs found by \cg, while in the latter they used \cg to help debugging after observing a failure. 

Another, very relevant subtree of \emph{software quality} is the one related to the automation of \emph{testing} activities. While we found an instance in which \cg was used to fix flaky tests \cite{pr:1395}, in the rest of cases (17 instances) the automated task is the generation of tests, \eg ``\emph{includes tests, which were coded with care by \cg}'' \cite{pr:1433}, ``\emph{tests were generated by \cg and while it was not perfect it did a decent job at creating the unit tests code}'' \cite{pr:1205}. \faWarning~The latter is only one of the comments we found in PRs which confirm the usefulness of \cg as an aid to write tests rather than as a completely automated solution: ``\emph{I have generated the tests with the help of \cg and manually checked all of them --- it got a few of them wrong or was testing impossible cases, but it did find that one edge case}''. \faLightbulbO~On this line, research approaches could aim at integrating LLM-based test generation with approaches aimed at identifying and repairing broken tests \cite{Choudhary11,DanielGM10,Pan22,Stocco18}.

\subsection{Documentation}
The last popular application of \cg we discuss (105 instances in our sample) concerns the automated generation of software \emph{documentation}. \cg is used both to write documentation from scratch (38 instances) as well as to improve existing documentation (see \emph{improving writing} category with 58 instances). In some cases, projects' users suggest to improve parts of the documentation since they found it difficult to read: ``\emph{I found the README.md a bit difficult to digest, so I utilized \cg to help me simplify the content. This allowed me to better understand the library's core features and functionality. It might be worth considering a shorter, more concise version of the README for easier comprehension}'' \cite{pr:1451}. 

\faLaptop~It could be useful for projects' owners, especially for non-native speakers of the language used in the documentation, to consider the usage of \cg to improve documentation quality. 

For what concerns the generation of documentation from scratch, \cg is mostly used for \emph{commenting code}, but also for drafting \emph{terms of service} \cite{pr:1463}, \emph{user guides} \cite{issue:643}, and \emph{README} files \cite{pr:1237}. Differently from what we found for tasks related to code generation, we did not observe negative reactions of projects' owners/reviewers. This is likely due to: (i) the excellent performance of the LLM when dealing with natural language; (ii) the higher likelihood that the contributor posting \cg-generated content has the actual competencies to assess whether the generated output is correct (\ie less coding skills required); (iii) the fact that, as \cg generates natural language, projects' contributors see pretty obvious (and relatively straightforward) ways to improve/adapt it when necessary, and therefore there is less evidence of complaints; and 
(iv) the lower risk related to errors in the generation task (\eg typos \emph{vs} bugs) except, of course, for terms-of-service. 
\faLightbulbO~An empirical investigation aimed at studying the sentiment of reviewers when inspecting different types of AI-generated contributions (\eg code \emph{vs} documentation) could help in better characterizing and backing up our observation. Last, but not least, also in this case, a proper (in some cases large) prompt may be needed by \cg to generate exhaustive and correct documentation. 

This, in turn, may stimulate research on how to combine LLMs with software reverse engineering approaches for that purpose.
	% !TEX root = main.tex
%%%%%%%%%%%%%%%%%%%%%%%%%%%%%%%%%%%%%%%%
%%%%%%%%%%%%%%%%%%%%%%%%%%%%%%%%%%%%%%%%
\section{Threats to Validity} \label{sec:threats}
%%%%%%%%%%%%%%%%%%%%%%%%%%%%%%%%%%%%%%%%
%%%%%%%%%%%%%%%%%%%%%%%%%%%%%%%%%%%%%%%%

Threats to \emph{construct validity} concern the relationship between the theory and observation. Studying the purpose of the use of \cg in software development by mining software repositories has an intrinsic limitation. This is because we observe only cases where developers mention \cg explicitly in a commit message, issue, or PR description. There could be other changes in which developers silently leveraged \cg.

Moreover, as explained in \secref{sec:design}, we analyzed the textual content of commit messages, issues, and PRs, as they could be queried by GitHub. However, there may be other places where \cg could have been mentioned, \eg code comments. These would require analyzing all projects' source code and could be considered in future work. 

A further threat is due to our interpretation of \cg purposes of usage, by reading and labeling commits and developers' discussions. This classification could have been affected by subjectiveness and imprecision. As explained in \secref{sec:design}, we mitigated this threat by having two annotators labeling each instance independently, and, after that, having a cooperative conflict resolution. 

Threats to \emph{internal validity} concern confounding factors internal to our study that could affect our results. During the manual analysis, we explicitly excluded cases in which the contribution of \cg to a given development activity was unclear. Also, we used multiple labels where \cg was used for multiple purposes.

Threats to \emph{external validity} concern the generalizabiity of our findings. Within the construct validity threats stated above, the observed findings limit to open-source projects hosted on GitHub only. 

Therefore, our study needs to be complemented by other types of studies (\eg interviews, survey questionnaires, ethnographic studies) conducted in closed-source scenarios, such as industrial environments. Moreover, we are only observing the first six months of \cg usage, and it is possible that its variety of use will largely increase in the future. Last, this study is only limited to \cg, and should be, in the future, extended to other general-purpose chat bots that could be used in software development, \eg including the recently-released Google Bard \cite{bard}. It is possible that some of them could adopt techniques to circumvent limitations/risks we found for \cg or, on the other hand, have limitations that \cg does not have, including the ability to access up-to-date content.
	% !TEX root = main.tex
%%%%%%%%%%%%%%%%%%%%%%%%%%%%%%%%%%%%%%%%
%%%%%%%%%%%%%%%%%%%%%%%%%%%%%%%%%%%%%%%%
\section{Related Work} \label{sec:related}
%%%%%%%%%%%%%%%%%%%%%%%%%%%%%%%%%%%%%%%%
%%%%%%%%%%%%%%%%%%%%%%%%%%%%%%%%%%%%%%%%

%Works related to the usage of LLMs in Software Engineering
This section describes related work about the use of LLMs-recom\-menders in software development activities. We focus  on studies (i) investigating the usage of ChatGPT when employed for software-related tasks; and (ii)  other state-of-the-art AI-based recommenders such as GitHub Copilot \cite{copilot}.  

\subsection{\cg for Software-related Tasks}
White \etal \cite {white2023chatgpt} present a catalog of prompt patterns aimed at exploiting \cg in three families of tasks, namely \emph{requirements elicitation}, \emph{system design and simulation}, \emph{code quality}, \emph{refactoring}. These families include a total of 14 sub-tasks (\eg \emph{requirements simulator}, \emph{code clustering}). 

While experimenting with different prompts, the authors show that, to date, a large human contribution is still needed to take advantage of LLMs for the automation of software-related tasks. 

Nascimento \etal \cite {nascimento2023comparing} compare the performance\footnote{With ``performance'' the authors mean the ability to properly accomplish a development task.} of software engineers with that of \cg as representative of an AI-based developer. They compared the code written by humans and by AI, finding that when it comes to simple programming tasks, \cg often outperforms novice programmers. Conversely, when the complexity of the tasks increases and more experienced programmers are needed to solve them, humans perform better than the AI.

Tian \etal \cite {tian2023chatgpt} investigate the potential of \cg when used to automate code generation, program repair, and code summarization. They compare \cg against state-of-the-art techniques showing that: (i) for code generation, \cg outperforms competitive techniques, while still struggling to generalize for new problems; (ii) in program repair, \cg achieves competitive performance when compared to the state-of-the-art technique; and (iii) when summarizing code, \cg does not always accurately explain the intention of a given code. Some of the limitations highlighted by Tian \etal (\eg the difficulties experienced by \cg when dealing with new likely unseen code elements) have been confirmed in our study.

Sridhara \etal \cite {sridhara2023chatgpt} evaluate the ability of \cg to solve software-related tasks such as method name suggestion, code review, and log summarization. Overall, fifteen different tasks have been considered, with \cg also in this case compared against the state-of-the-art and/or against human expert ground truths. 

Their results show that for most tasks the answers provided by \cg are credible and sometimes better than the state-of-the-art or human expert output. However, \cg poorly performs in the tasks of vulnerability detection and test prioritization.

Dong \etal \cite {dong2023self} set up an experimental design aimed at mimicking human teamwork via \cg: The authors assemble a virtual team composed of three \cg instances instructed to cover different roles, namely analyst, programmer, and tester. The goal of the team is to carry out software analysis, coding, and testing. The experimental results show that the code generated by this ``virtual'' collaboration outperforms 
%self-collaborating virtual relatively improves 29.9\%-47.1\% Pass@1 compared to 
direct code generation. Moreover, the self-collaboration allows \cg to address more complex tasks as compared to those achievable via direct code generation.

Our study complements the work discussed above and provides empirical knowledge about the use of \cg in the open source. This, on the one hand, stimulates further empirical research focused on specific tasks that have been found relevant for developers. On the other hand, stimulates research on approaches to circumvent the clear limitations experienced by developers when using \cg.

Several researchers investigated the usage of \cg for automated program repair (APR). Cao \etal \cite {cao2023study} focus on program repair in the context of deep learning programs, which are known to pose specific challenges for their debugging and testing. 
In particular, the authors analyze the capability of \cg to identify faulty programs, localize the faults, and automatically repair it. Also, they show how different prompts/dialogues can have a major impact on the \cg capabilities of addressing the tackled tasks.

Sobania \etal \cite {sobania2023analysis} assess the capability of \cg to fix bugs in the QuixBugs benchmark. \cg performances have been contrasted against the state-of-the-art techniques, showing its competitiveness against both traditional and deep learning-based APR techniques.

Fan \etal \cite {fan2022automated} present a study related to the previously discussed ones, but assess whether APR techniques can be used to fix erroneous solutions (code) generated by LLMs. They found that the code automatically generated by LLMs share common mistakes also present in human-written code, suggesting that APR techniques can be suitable for automated code as well. 

These pieces of research confirm the potential of \cg for APR that, accordingly to our taxonomy, is a task for which it has been largely leveraged by developers.

%Finally, Jalil \etal \cite{JalilRLML23} investigated the promises and perils of using \cg for software testing education. Specifically, they asked \cg to answer quiz or solve simple problems taken from software testing textbooks.   \cg was able to answer 77.5\% of the questions correctly. Moreover, it provided fairly correct answers in 55.6\% of the cases, and was able to add further correct explanations for 53.0\% of the cases.

\subsection{Empirical Studies on GitHub Copilot}

Several studies have analyzed the use of GitHub Copilot \cite{copilot}. These investigations include studies on the impact of the recommender on developers \cite{imai2022github, peng2023impact, vaithilingam2022expectation, ziegler:maps2022}, robustness assessment \cite{mastropaolo2023robustness, yetistiren2022assessing, wong2022exploring}, empirical evaluation of correctness \cite{nguyen2022empirical, yetistiren2022assessing}, and scrutiny of security aspects \cite{pearce2021empirical, sobania2021choose, asare2022github}.

% Effectiveness
Imai \etal \cite{imai2022github} investigate the extent to which Copilot is a valid alternative to a human pair programmer. The authors involved 21 participants each of which performed coding tasks under three different treatments: (i)  pair-programming with Copilot; (ii) human pair-programming as a driver; and (iii); human pair-programming as a navigator. They observed that Copilot results in increased productivity (\ie number of added lines of code), but decreased quality in the produced code.

Peng \etal \cite{peng2023impact} achieved outcomes consistent with the findings of Imai \etal \cite{imai2022github}. In particular, they found that the group of developers with access to the AI pair programmer (\ie treatment group), completed the task 55.8\% faster than the control group, \ie those who did not use the code recommender. 

Vaithilingam \etal \cite{vaithilingam2022expectation} observed instead that Copilot does not improve the task completion time and success rate. However, developers report that they value Copilot's support since it can recommend code that can be used as a starting point for the task, thus saving the effort of searching online.

%Productivity Assessment of Neural Code Completion
Ziegler \etal \cite{ziegler:maps2022} examine the relationship between Copilot usage  and developers' productivity. Their findings indicate that the acceptance rate of suggested solutions can act as a reliable proxy for perceived productivity.

%An Empirical Evaluation of GitHub copilot's Code Suggestions
Nguyen and Nadi \cite{nguyen2022empirical} provide LeetCode questions as input to Copilot to assess its ability to provide correct solutions. The study revealed significant variations in correctness among the questions related to different programming languages, ranging from 57\% for Java down to 27\% for JavaScript.

%\eject

Yetistiren \etal \cite{yetistiren2022assessing} set out to evaluate the quality of code generated by GitHub Copilot. The assessment included aspects such as validity, correctness, and efficiency of the generated code. In the reported study GitHub Copilot achieved a remarkable 91.5\% success rate in generating syntactically valid code. Further analyses revealed that 28.7\% of the completed tasks were actually correct,  51.2\% were partially correct, and 20.1\% were wrong.

Wong \etal \cite{wong2022exploring} present an investigation exploring the extent to which Copilot is capable of generating formally verifiable code. Their findings reveal that the deep learning-based code recommender successfully synthesized formally verifiable code for four out of the six problems subject of the experiment.

Mastropaolo \etal \cite{mastropaolo2023robustness} study the robustness of GitHub Copilot when used for the automated generation of Java methods. They provided Copilot with different but semantically equivalent descriptions of the methods to generate, showing that the provided prompt can substantially change the generated recommendation. In particular, this was observed for $\sim$46\% of methods involved in the empirical study.

%Security
Hammond \etal \cite{pearce2021empirical} assess the likelihood of receiving code recommendations from Copilot that feature security flaws. Their findings indicate that 40\% of the experimented completion scenarios led to the injection of vulnerable code. 
In a related study, Asare \etal \cite{asare2022github} investigate whether the Copilot increases the likelihood of introducing vulnerabilities as compared to manual coding. To this aim, the authors prompted Copilot to generate code recommendations in contexts in which human developers had introduced vulnerabilities in the past. The study revealed that Copilot also produced original vulnerable code in $\sim$33\% of cases.

Sobania \etal \cite{sobania2021choose} compare the program synthesis capabilities of  GitHub Copilot with those of genetic programming techniques. They conclude that the two approaches have similar performance.

The studies discussed above mostly concern code generation, yet recently Copilot is also being used to generate other artifacts such as commit messages. 

Our study, conducted with a more generic tool such as \cg, indicates that code generation is only one task where developers may leverage AI-based tools.
	
	% !TEX root = main.tex
%%%%%%%%%%%%%%%%%%%%%%%%%%%%%%%%%%%%%%%%
%%%%%%%%%%%%%%%%%%%%%%%%%%%%%%%%%%%%%%%%
\definecolor{lightlightgrey}{RGB}{233, 236, 239}

\begin{table*}
\centering
\caption{Summary of implications for practitioners and researchers derived from our study}\vspace{-0.3cm}
\label{tab:implications}
{\footnotesize
\resizebox{.9\textwidth}{!}{%
\begin{tabular}{p{16cm}}

\bottomrule
\rule{0pt}{4ex} \cellcolor{black} \textcolor{white}{\large \faLaptop \hspace{2 mm}  \bf \small Insights for Practitioners}  \\[2ex] \hline
\rule{0pt}{3ex} \cellcolor{lightlightgrey} \textbf{\small Contributions including AI-generated content}
\vspace{0.04cm}\\[1ex]

\begin{itemize}[leftmargin=0.5cm]
\setlength\itemsep{0.5em}
\item Define guidelines for projects' contributions including AI-generated code, \eg a project may decide to only welcome AI-generated code from users that are confident in assessing the correctness of the contributed code.

\item Clear risk related to the ownership and understanding of code contributed via \cg, especially when it is used to contribute with a complete feature: The (human) contributor is not always able to explain or advocate for the submitted code.

\item As with any AI-based solution, the usage of \cg for software-related tasks may result in artificial hallucination: AI responses that look plausible to the user can be clearly wrong. The hard skills of developers remain essential in the era of AI-assisted coding.
\end{itemize}\vspace{0.01cm}\\\hline

%%%%%%%%%%%%%%%%%%%%%%%%%%%%%%%%%%%%%%%%%%%%%%%%%%%%%%%%%%%%%%%%%%%%%%%%%%%
%%%%%%%%%%%%%%%%%%%%%%%%%%%%%%%%%%%%%%%%%%%%%%%%%%%%%%%%%%%%%%%%%%%%%%%%%%%
%%%%%%%%%%%%%%%%%%%%%%%%%%%%%%%%%%%%%%%%%%%%%%%%%%%%%%%%%%%%%%%%%%%%%%%%%%%

\rule{0pt}{3ex}  \cellcolor{lightlightgrey} \textbf{\small Automation possibilities offered by \cg}
\vspace{0.04cm}\\[1ex]

\vspace{0.04cm}\cg can be leveraged to support very complex tasks, for which its usage has not been documented/experimented in the literature. These include:\vspace{0.1cm}

\begin{itemize}[leftmargin=0.5cm]
\setlength\itemsep{0.5em}
\item Prototyping the complete first version of a project, providing a substantial jumpstart in software development.

\item TDD collaboration, where the developer is mostly in charge of writing tests and delegating to LLM the code writing task.

\item Translating source code across different programming languages, thus improving code reusability.

\item Release planning, suggesting ideas on how to improve a software project based on what was observed in the wild.

\item Data generation, \eg augmenting UI-related strings handling dialogs with the user.

\item Debugging, from several different perspectives, including helping in locating the bug as well as in reproducing it.
\end{itemize}\vspace{0.01cm}\\\hline

%%%%%%%%%%%%%%%%%%%%%%%%%%%%%%%%%%%%%%%%%%%%%%%%%%%%%%%%%%%%%%%%%%%%%%%%%%%
%%%%%%%%%%%%%%%%%%%%%%%%%%%%%%%%%%%%%%%%%%%%%%%%%%%%%%%%%%%%%%%%%%%%%%%%%%%
%%%%%%%%%%%%%%%%%%%%%%%%%%%%%%%%%%%%%%%%%%%%%%%%%%%%%%%%%%%%%%%%%%%%%%%%%%%

\rule{0pt}{3ex} \cellcolor{lightlightgrey} \textbf{\small Software-related tasks involving natural language}
\vspace{0.04cm}\\[1ex]

\vspace{0.04cm} Due to its extensive training on natural language artifacts, \cg is well-suited to support software-related tasks strongly characterized by natural language, such as the generation of software documentation. %This is likely due to the excellent performance of LLMs when dealing with natural language as compared to source code.
\vspace{0.4cm}\\\hline

%%%%%%%%%%%%%%%%%%%%%%%%%%%%%%%%%%%%%%%%%%%%%%%%%%%%%%%%%%%%%%%%%%%%%%%%%%%
%%%%%%%%%%%%%%%%%%%%%%%%%%%%%%%%%%%%%%%%%%%%%%%%%%%%%%%%%%%%%%%%%%%%%%%%%%%
%%%%%%%%%%%%%%%%%%%%%%%%%%%%%%%%%%%%%%%%%%%%%%%%%%%%%%%%%%%%%%%%%%%%%%%%%%%

\rule{0pt}{3ex} \cellcolor{lightlightgrey} \textbf{\small Risks related to sensible/private information}
\vspace{0.04cm}\\[1ex]

\vspace{0.04cm}Some of the tasks automated via \cg (\eg code review) require to pass it sensible information, such as the code base itself, which may not be acceptable in industrial environments. Practitioners must carefully consider the tradeoff of using a publicly available LLM \emph{vs} training a local LLM.\vspace{0.4cm}\\\hline

%%%%%%%%%%%%%%%%%%%%%%%%%%%%%%%%%%%%%%%%%%%%%%%%%%%%%%%%%%%%%%%%%%%%%%%%%%%
%%%%%%%%%%%%%%%%%%%%%%%%%%%%%%%%%%%%%%%%%%%%%%%%%%%%%%%%%%%%%%%%%%%%%%%%%%%
%%%%%%%%%%%%%%%%%%%%%%%%%%%%%%%%%%%%%%%%%%%%%%%%%%%%%%%%%%%%%%%%%%%%%%%%%%%

\rule{0pt}{3ex} \cellcolor{lightlightgrey} \textbf{\small Unsuitability of \cg for tasks dealing with recent technologies}
\vspace{0.04cm}\\[1ex]

\vspace{0.04cm}\cg may not be suitable for tasks requiring up-to-date technology appeared after its last retraining. LLMs leveraging up-to-date knowledge available in the wild may obtain better results.  %sources of information (\eg recently presented frameworks, programming languages, \etc), since they are usually not frequently retrained and likely to make up wrong answers in this scenario.
\vspace{0.3cm}\\\hline

%%%%%%%%%%%%%%%%%%%%%%%%%%%%%%%%%%%%%%%%%%%%%%%%%%%%%%%%%%%%%%%%%%%%%%%%%%%
%%%%%%%%%%%%%%%%%%%%%%%%%%%%%%%%%%%%%%%%%%%%%%%%%%%%%%%%%%%%%%%%%%%%%%%%%%%
%%%%%%%%%%%%%%%%%%%%%%%%%%%%%%%%%%%%%%%%%%%%%%%%%%%%%%%%%%%%%%%%%%%%%%%%%%%

\rule{0pt}{4ex} \cellcolor{black} \textcolor{white}{\large \faLightbulbO \hspace{2 mm}  \bf \small Insights for Researchers}  \\[2ex] \hline
\rule{0pt}{3ex} \cellcolor{lightlightgrey} \textbf{\small Implications for the design of empirical studies}
\vspace{0.04cm}\\[1ex]

\begin{itemize}[leftmargin=0.5cm]
\setlength\itemsep{0.5em}
\item Empirical investigations studying OSS contributors  may or may not consider representative developers that only submitted AI-generated code.

\item \cg must be considered as a baseline in works proposing novel recommenders for tasks where it was found to be useful. However, as the dataset on which \cg has been trained is not publicly available, it is hard to make a fair comparison ensuring the lack of overlap between training and test set. A possible solution is to use recent data points as test set, since those are unlikely to have been seen by the model behind \cg. %For example, for commit message generation, this may mean using the 10\% most recent commits as a test set.
\end{itemize}\vspace{0.01cm}\\\hline

%%%%%%%%%%%%%%%%%%%%%%%%%%%%%%%%%%%%%%%%%%%%%%%%%%%%%%%%%%%%%%%%%%%%%%%%%%%
%%%%%%%%%%%%%%%%%%%%%%%%%%%%%%%%%%%%%%%%%%%%%%%%%%%%%%%%%%%%%%%%%%%%%%%%%%%
%%%%%%%%%%%%%%%%%%%%%%%%%%%%%%%%%%%%%%%%%%%%%%%%%%%%%%%%%%%%%%%%%%%%%%%%%%%

\rule{0pt}{3ex} \cellcolor{lightlightgrey} \textbf{\small Studying and enhancing AI-aided development processes}
\vspace{0.04cm}\\[1ex]

\vspace{0.04cm}Practitioners are already leveraging \cg for a variety of tasks. Nevertheless, it may be useful to (empirically) devise AI-enabled development processes, with suitable guidelines. 
%some of them require empirical investigations to assess the extent to which LLMs actually help in boosting performance. 
These include using \cg (or similar tools):
\vspace{0.1cm}

\begin{itemize}[leftmargin=0.5cm]
\setlength\itemsep{0.4em}
\item In TDD, with the developer being mostly in charge of writing tests and delegating to the LLM the production code.
\item To support program comprehension, especially when newcomers onboard a project and must become familiar with its code base.
\item To generate tests.
\item To automate code review.
\end{itemize}\\

Moreover, researchers should focus on approaches aimed at better integrating \cg or similar tools in development contexts where there is a need for:
\begin{itemize}[leftmargin=0.5cm]
	\setlength\itemsep{0.4em}
\item Understand, refactor, complete very specific code or other artifacts.
\item Avoid exposing internal artifacts to the outside, \eg using Retrieval Augmentation Generation or similar approaches.
\item Repair \cg-generated code and tests, or adapt them in own code base.
\end{itemize}\vspace{0.01cm}\\\hline
%%%%%%%%%%%%%%%%%%%%%%%%%%%%%%%%%%%%%%%%%%%%%%%%%%%%%%%%%%%%%%%%%%%%%%%%%%%
%%%%%%%%%%%%%%%%%%%%%%%%%%%%%%%%%%%%%%%%%%%%%%%%%%%%%%%%%%%%%%%%%%%%%%%%%%%
%%%%%%%%%%%%%%%%%%%%%%%%%%%%%%%%%%%%%%%%%%%%%%%%%%%%%%%%%%%%%%%%%%%%%%%%%%%

\rule{0pt}{3ex} \cellcolor{lightlightgrey} \textbf{\small Questioning the suitability of existing recommenders for AI-generated code}
\vspace{0.04cm}\\[1ex]

\vspace{0.04cm}The effectiveness of recommender systems for software engineers proposed in the literature (\eg tools identifying and fixing internationalization issues, APR techniques) may need to be reassessed on AI-generated code, since the latter may have characteristics different from those of human-written code.\vspace{0.04cm}\\\hline

\end{tabular}
}
}

\end{table*}

\section{Conclusions} \label{sec:conclusion}
%%%%%%%%%%%%%%%%%%%%%%%%%%%%%%%%%%%%%%%%
%%%%%%%%%%%%%%%%%%%%%%%%%%%%%%%%%%%%%%%%

In this paper, we manually analyzed, through an open coding process, \manually commits, issues and PRs from open source projects in which there was documented usage of \cg for the automation of software-related tasks. The goal was to categorize the type of support \cg provided. 
The result of this analysis is a taxonomy of \categories tasks (partially) automated via \cg, which we discussed highlighting \cg's strengths and weaknesses and distilling implications for practitioners and researchers. The latter, together with our taxonomy, represent the main outcome of our study, and have been abstracted and summarized in \tabref{tab:implications} for easier reference. Our future work will focus on validating our findings by (i) interviewing developers, and (ii) generalizing them to other general-purpose LLMs.

	In addition, we plan to obtain evidence of the degree to which \cg  has proven beneficial for developers in the context of software-related task. Achieving this, however, requires the implementation of a meticulous study design with the explicit aim of mitigating biases and addressing significant issues that may arise during the analysis. We deliberately opted against undertaking this investigation by purely mining software repositories, as developers may be less prone to report cases in which usage attempts of \cg turned out to be a failure.

	\section*{Acknowledgment}
	This project has received funding from the European Research Council (ERC) under the European Union's Horizon 2020 research and innovation programme (grant agreement No. 851720). Massimiliano Di Penta acknowledges the Italian PRIN 2020 Project EMELIOT ``Engineered MachinE Learning-intensive IoT system'',  ID 2020W3A5FY. Federica Pepe is partially funded by the PNRR DM 352/2022 Italian Grant for Ph.D. scholarships. 
	
	Tufano thanks CHOOSE for sponsoring her trip to the conference.

	\bibliographystyle{ACM-Reference-Format}
	\bibliography{main}
	
	\balance
	
\end{document}